\documentclass[12pt,preprint]{aastex}

\shorttitle{Metallicities, ages and kinematics in $\omega$~Cen}
\shortauthors{Sollima et al.}

\begin{document}

\title{Metallicities, relative ages and kinematics of stellar populations in
$\omega$~Centauri\footnote{Based on FLAMES/GIRAFFE observations collected with the Very Large 
Telescope at the European Southern Observatory, Cerro Paranal, Chile, 
within the observing programs 71.D-0217(A) and 74.D-0369(A).}}


\author{A. Sollima\altaffilmark{1}, E. Pancino\altaffilmark{2}, F. R.
Ferraro\altaffilmark{1}, M. Bellazzini\altaffilmark{2}, O. Straniero\altaffilmark{3} and L. Pasquini\altaffilmark{4}}

\altaffiltext{1}{Dipartimento di Astronomia, Universit\`a di Bologna, via Ranzani 1, I-40127 Bologna, Italy}
\altaffiltext{2}{Osservatorio Astronomico di Bologna, via Ranzani 1, I-40127 Bologna, Italy}
\altaffiltext{3}{Osservatorio Astronomico di Collurania, via M. Maggioni, I-64100 Teramo, Italy}
\altaffiltext{4}{European Southern Observatory, Karl-Schwarzchild-Strasse 2, D-85748 Garching bei Munchen, Germany}


\begin{abstract}
We present results of an extensive spectroscopic
survey of Subgiant stars in the stellar system $\omega$ Centauri. Using infrared
Ca II triplet lines, we derived metallicities and radial velocities for more
than 250 stars belonging to different stellar populations of the system.  
We find that the most metal rich component, the anomalous Sub Giant Branch
(SGB-a), has a metallicity of 
[Fe/H] $\sim$ -0.6 fully compatible with that determined along the 
anomalous Red Giant Branch (RGB-a). 
Our analysis suggests that the age of this component and of the other
metal-intermediate ($-1.4~<~[Fe/H]~<~-1.0$) stellar populations of the system 
are all comparable to that of the
dominant metal poor population within 2 Gyr, regardless of any choice of helium abundance.
These results impose severe constraints on the
time-scale of the enrichment process of this stellar system, excluding
the possibility of an extended star formation period.
The radial velocity analysis of the entire sample demonstrates
that only metal-intermediate populations are kinematically cooler than the
others.
\end{abstract}


\section{Introduction}

The understanding of the origin and star formation history of the stellar system 
$\omega$ Centauri (NGC 5139), the
most massive and luminous globular cluster of the Milky Way ($M \sim 2.9 ~ 10^6 M_{\sun}$, 
Merrit et al. 1997), represents one of the most intriguing brain teasers
of stellar astrophysics. The observational evidences collected over the last 40
years indicates that $\omega$ Cen is the most peculiar object among galactic
star clusters in terms of structure, kinematics and stellar content. It is the
only known globular cluster which shows a clear metallicity spread.
In the last ten years, extensive spectroscopic
surveys have been performed on large samples of giant stars (Norris, Freeman \&
Mighell 1996, Suntzeff \& Kraft 1996, hereafter NFM96 and SK96 respectively),
showing a multimodal distribution of heavy-element.

Recent photometric surveys have revealed the presence of several anomalous
sequences in the color-magnitude diagram (CMD). In particular, wide-field photometric studies (Lee et al.
1999, Pancino et al. 2000 -- hereafter P00) have shown the presence of an
additional anomalous Red Giant Branch (hereafter RGB-a). According to P00, this population contains
approximately 5\% of the cluster's stellar content and represents the extreme
metal-rich end of the metallicity distribution. Moreover, the
metal-rich giants appear to have different spatial distribution and dynamical
behaviour with respect to the metal poor ones (Norris et al. 1997, Jurcsik
1998, Pancino et al. 2000, 2003, Sollima et al. 2005a). Beside the dominant
population (having metallicity $[Fe/H] \sim -1.6$) and the RGB-a ($[Fe/H] \sim
-0.6$), three metal-intermediate (MInts) components (with $-1.3 < [Fe/H] <
-1.0$) have been identified by Sollima et al. (2005a, hereafter S05). More
recently, a high-precision photometric survey in the central region of the
cluster (Ferraro et al. 2004, hereafter F04) revealed the presence of a narrow 
well defined Sub Giant Branch (SGB-a) which merges into the Main Sequence (MS) 
of the dominant cluster population at a magnitude significantly fainter than
the cluster Turn-off (TO). Although this feature seems to be the extension of
the RGB-a, a direct comparison of the MS-TO morphology with theoretical
isochrones shows that the anomalous metal-rich population cannot be
significantly younger than the most metal-poor one. This interpretation leads to
inconsistences with the self-enrichment scenario proposed for $\omega$ Cen
(Freeman 1993, Lee et al. 1999, Bekki \& Freeman 2003). Finally, Anderson
(2002) and Bedin et al.
(2004) discovered new peculiarities also along the MS of the
cluster. In fact,
an additional less populous blue Main Sequence (bMS, comprising $\sim$ 30\% of
the whole cluster MS stars) located parallel to the
dominant one, has been evidenced. 
According to stellar models with canonical chemical abundances, the location
of the observed bMS would suggest a lower metallicity. 
Conversely, Piotto et al. (2005,
hereafter P05) presented the spectroscopical analysis of 34 MS stars belonging
to the two MS components showing that the bMS stars present a
metallicity $\sim 0.3~dex$ higher than the dominant cluster population. This
evidence brought P05 to conclude that only a large helium enhancement could
explain the anomalous position of the bMS in the CMD, as
suggested by Norris (2004, hereafter N04) and Lee et al. (2005).   

Until now, only rough estimates of the relative ages of $\omega$ Cen have been made,
using broad and narrow band photometry (Hilker \& Richtler 2000, Rey et al.
2004, Hughes et al. 2004). These studies suggested a time scale of chemical
enrichment up to 6 Gyr. A smaller age spread ($\sim$ 2 Gyr) has been
estimated by N04, Hilker et al. (2004) and Lee et al. (2005), assuming different
helium abundances among the stellar populations of $\omega$ Cen. In particular,
Hilker et al. (2004) estimated an age difference  $\Delta t \sim 1.6~Gyr$
between the MP and MInt populations by measuring iron abundances of $\sim 400$
SGB-TO stars with low resolution line indexes.

As part of a long term project devoted to the detailed study of the properties
of different stellar populations in this cluster (see Ferraro et al. 2003), we
present low-resolution spectroscopy of 256 SGB stars of $\omega$ Cen. We
derived metal content and radial velocities for the different sub-populations
of the cluster. Using these estimates we derived their relative ages
by the comparison with theoretical isochrones. 

In \S 2 we describe the observations and the 
data reduction techniques. In \S 3 we present the metallicity distribution for
the whole sample. \S 4 is devoted to the estimation of the relative ages of the
observed populations. In \S 5 we show the dynamical behaviour of 
the populations of $\omega$ Cen as a function of metallicity. Finally, we 
summarize and discuss our results in \S 6.

\section{Observations and data reduction}

\begin{figure}
%\plotone{f1.eps}
\plotone{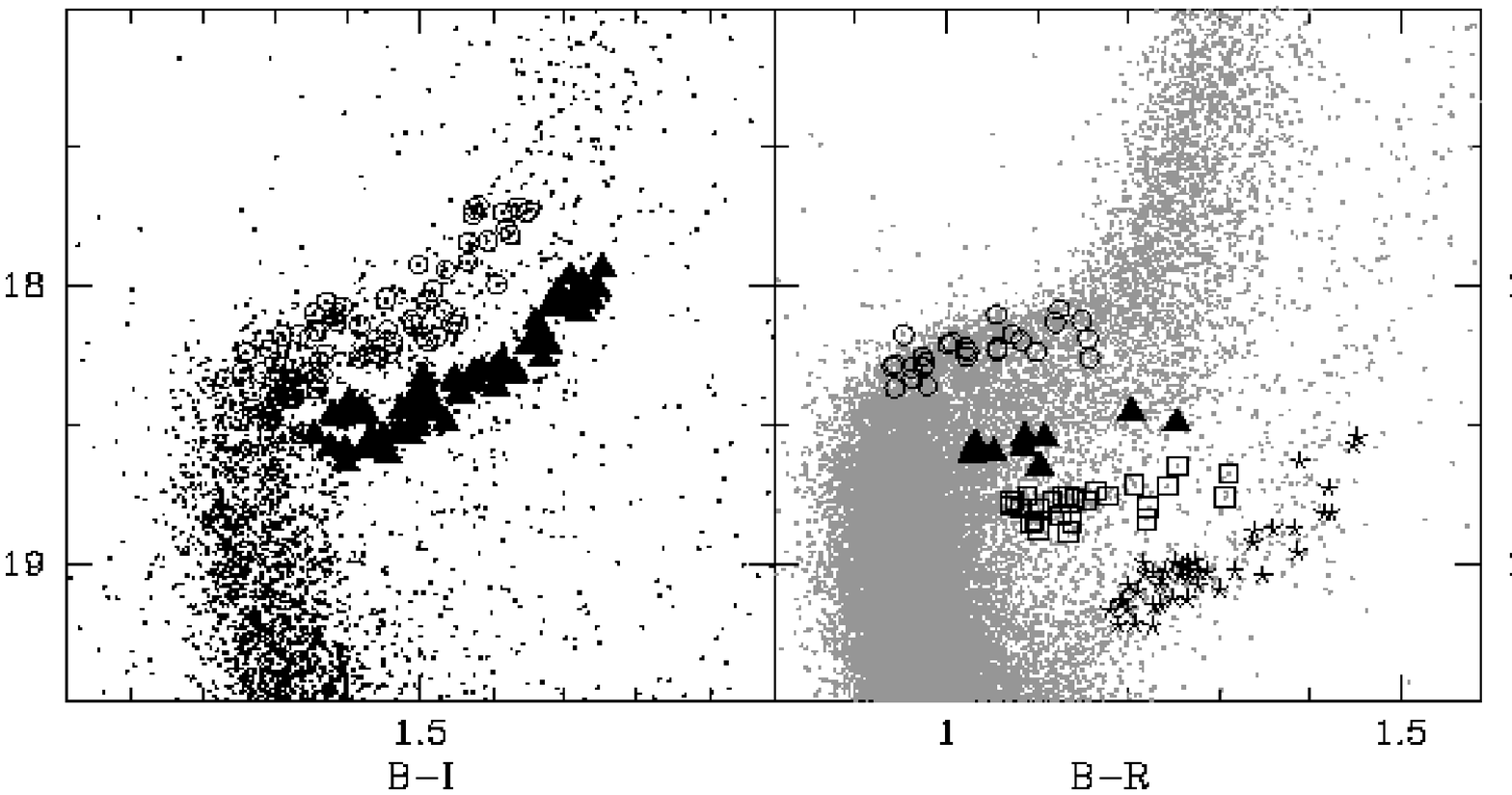}
\caption{The positions of the selected targets are marked in the CMDs of
$\omega$ Centauri. Stars selected from WFI
photometry by P00 are shown in the left $panel$, stars from ACS/HST 
observations (F04) are marked in the right $panel$. Different symbols indicate
the stars selection along different branches: SGB-MP (open circles), SGB-MInt2 
(filled triangles), SGB-MInt3 (open squares) and SGB-a (asterisks).} 
\label{sel}
\end{figure}

Observations were performed during two runs on May 2003, as a part of the 
Ital-FLAMES Guaranteed Time Observations (GTO), and on February 2005 at the VLT/UT2 at ESO (Cerro Paranal, 
Chile) equiped with the multi-fiber spectrograph FLAMES/MEDUSA (Pasquini et al.,
2003). We used the low resolution grating LR8, which allows a 
spectral coverage of 1200~\AA (between 8200-9400~\AA) with a resolving power 
of $R \sim 6500$. The spectra were obtained combining three
2085 s long exposures secured in good seeing conditions ($FWHM < 0.8"$) ,
reaching an average signal to noise ratio of $S/N \sim 50$ per pixel. We
observed two samples of data: {\it (i) the WFI sample:} 170
SGB stars belonging to two different branches detected in the CMD of P00
located in an external region of $\omega$ Cen ($\sim$ 10' away from the
cluster centre), and {\it (ii) the ACS sample:} 110 SGB stars, selected from the ACS@HST photometry by
F04, along four well separated branches. The target
stars selected on the CMDs are indicated in Fig. \ref{sel}. 

\begin{figure}
%\plotone{f2.eps}
\plotone{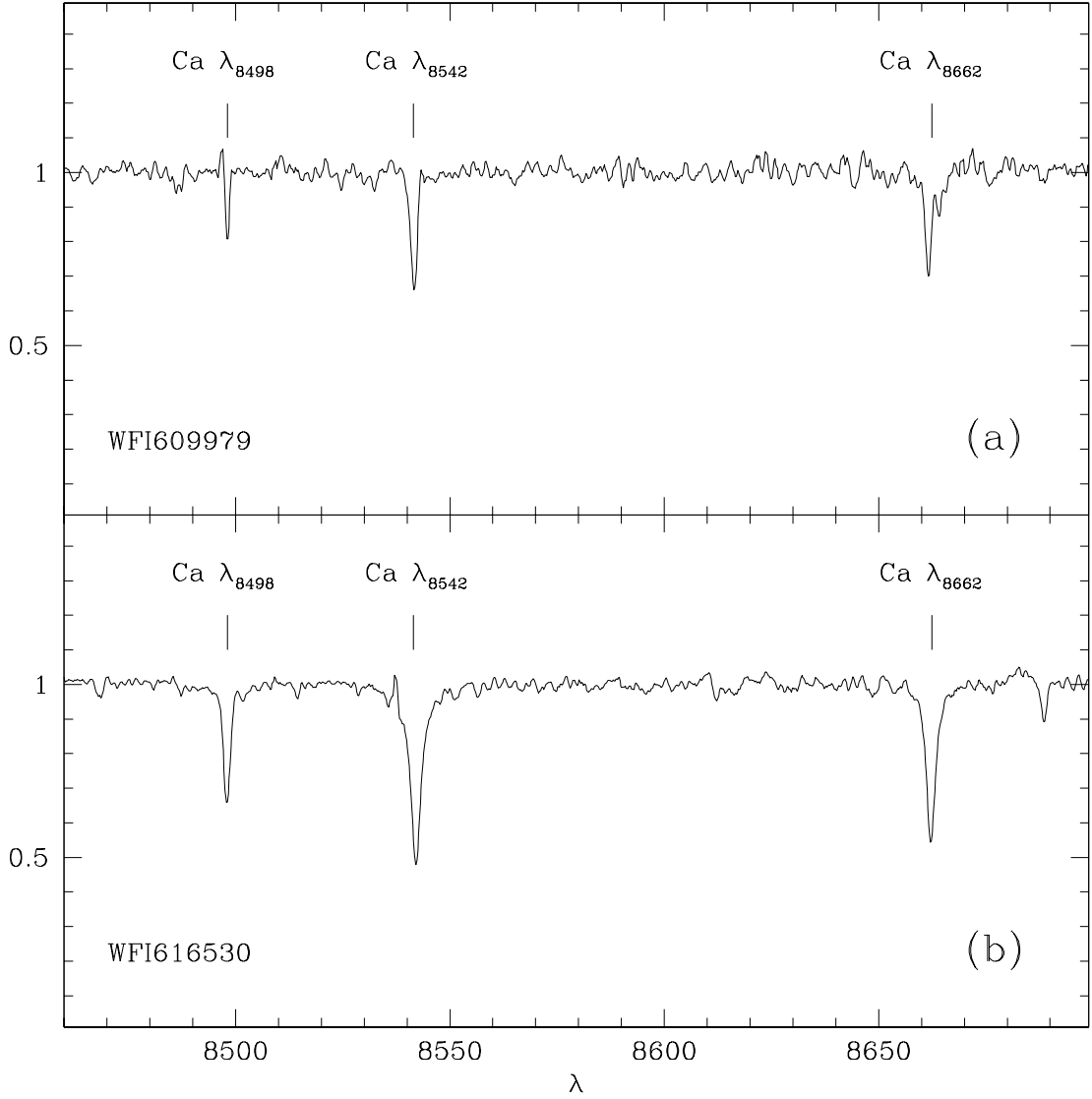}
\caption{Spectra of star WFI609979 ([Fe/H]=-2.15, $panel$ a) and WFI616530
([Fe/H]=-0.97, $panel$ b). Ca II triplet lines are indicated.} 
\label{templ}
\end{figure}

The one-dimensional spectra were extracted with the GIRAFFE pipeline. 16 fibers
were dedicated to sky observations in each exposure. An average sky
spectrum was obtained and subtracted from the object spectra, by taking into
account the different fiber transmission. The spectra were then
continuum normalized and corrected for telluric absorption bands with IRAF.
Spectra of two program stars are shown in Fig. \ref{templ} to illustrate the
quality of our data. The determination of equivalent widths (EWs) and radial
velocities was performed on the basis of infrared Ca II triplet lines strength,
following the prescriptions of Rutledge et al. (1997, hereafter R97).

\subsection{Radial Velocities}

In order to determine radial velocities, we cross correlated the spectra of our
sample with a high signal to noise ($S/N > 100$) RGB template spectrum (ROA 371,
see Pancino et al. 2002, hereafter P02). All spectra were corrected for
heliocentric velocity. The correlation function
$C(\lambda)$ between the program spectra $P(\lambda)$ and the template spectrum
$T(\lambda)$ was calculated as follows
$$
C(\lambda)=\sum_{\lambda=\lambda_{templ}^{inf}}^{\lambda_{templ}^{sup}}
P(\lambda)T(\lambda-\Delta \lambda)
$$
with $\lambda_{templ}^{inf}=8350~\AA$ and $\lambda_{templ}^{sup}=8750~\AA$ for
a large range of $\Delta \lambda$ values ($\pm$10\AA). The correlation function
$C(\lambda)$ was then fit with a gaussian function whose center provided the
velocity estimate. This procedure was performed independently on each single
exposure and the root mean square (rms) computed from the mean was assumed as the velocity error. 
The final $\Delta v_r$ estimate was used to doppler-shift the spectra at the 
wavelength corresponding to $v_r=0$, 
so that the band windows used to
calculate the EWs were properly aligned. All stars with
radial velocity different from the mean cluster velocity by more than $2 \sigma$
were considered field stars. Adopting this criterion, 256 out of a total of 280
stars were retained as bona fide cluster members.  

\subsection{Ca Index measurement}

We measured the EW of three Ca II lines and defined an {\it equivalent line
strength index} ($\Sigma~Ca$) according to R97: 
$$ \Sigma~Ca = 0.5 W_{8498} + W_{8542} + 0.6 W_{8662} $$  
Where $W_{8498}, W_{8542}~\mbox{and}~W_{8662}$ are the EWs of the
CaII lines at $8498 \AA, 8542 \AA ~\mbox{and}~8662 \AA$ respectively. 
The measurement of the EW of each Ca II line was done by linearly
interpolating the average intensities in two continuum bands on each side of the
feature. The EW was the integral over the line band of the difference between
the continuum and the line. The band limits used to define the continuum and the
feature regions are listed in Table 1. The line profile was modeled with a
Moffat function of exponent 2.5 as indicated by R97. For each program star, the
three exposures were averaged in order to obtain a higher signal to noise
spectrum. The index measurement was performed independently on each single
exposure and on the averaged spectrum. We adopted the $\Sigma~Ca$ index derived from
the averaged spectrum as reference, and the rms computed from
the three exposures has been assumed as $\Sigma~Ca$ error.

In all previous Ca II triplet analysis performed on globular cluster RGB
stars, the obtained $\Sigma~Ca$ indices were converted into "reduced" EW using
an empirical linear relation linking the $\Sigma~Ca$ index with the V magnitude
difference between the program star and the horizontal branch (HB) of the cluster
(Armandroff \& Da Costa 1991, Da Costa \& Armandroff 1995, Geisler 1995, R97,
Cole et al. 2004). This correction removes the gravity effect on EW that, for
giant stars, increases approximately linearly with magnitude, with a
negligible dependence on temperature (Idiart et al. 1997). Unfortunately, as
gravity increases, this approximation holds no more (SK96, Cenarro et al. 2001
-- hereafter C01). For this reason we decided to re-calibrate the relation
linking $\Sigma~Ca$ with metallicity, using the most extensive catalog of
observed spectra available in the literature by C01. A detailed description of
the calibration procedure is provided in Appendix~A.

\section{Metallicity distribution}

Fig. \ref{histo} shows the metallicity distribution obtained for the 152 member
stars belonging to the $WFI~sample$. A sharp and asymmetric peak
at [Fe/H] $\sim$--~1.7 can be seen, with a long tail extending toward higher
metallicity ([Fe/H] $>~-1.4$), in agreement with the previous spectroscopical
determinations by NFM96 and SK96. Care must be taken when comparing the
metallicity distribution shown in Fig. \ref{histo} with those obtained by NFM96
and SK96, since the present analysis is based on a {\it biased} sample, formed by
targets selected along the different SGBs of $\omega$ Cen. Nevertheless, the
global shape of the distribution in Fig. \ref{histo}, is consistent with
those showed by NFM96 and SK96. 

\begin{figure}
%\plotone{f4.eps}
\plotone{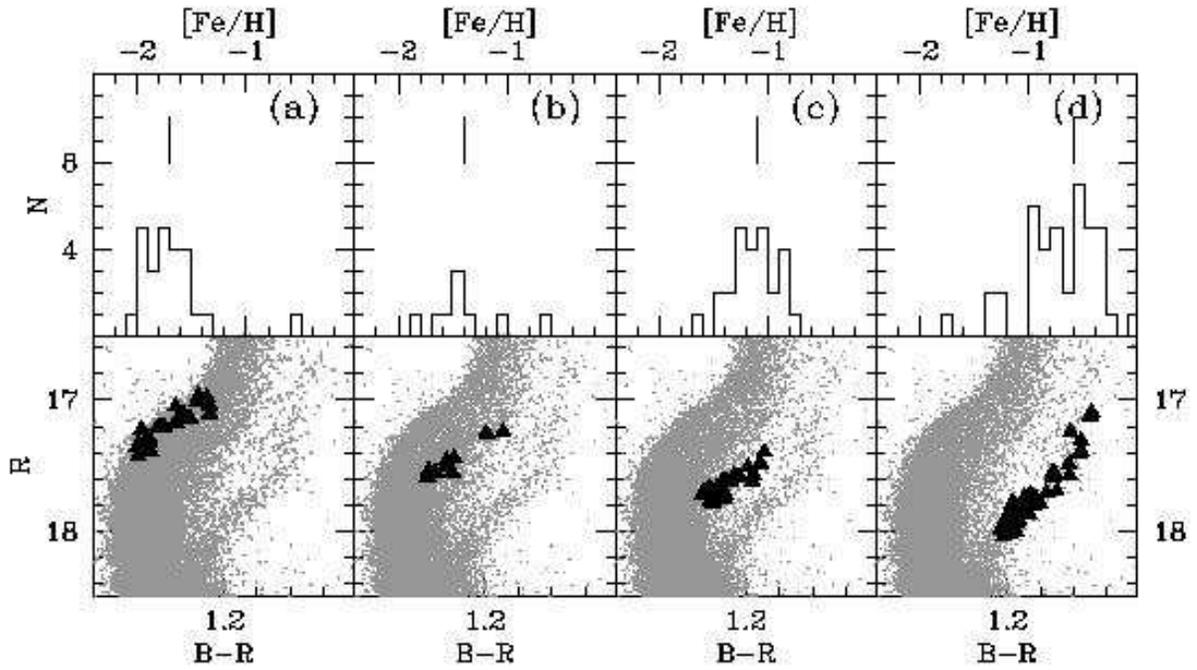}
\caption{Metallicity distributions for the four group of stars selected in the 
$ACS~sample$ (upper $panels$). The peak metallicity of each group is indicated. 
The positions of the selected stars are marked on
the ACS CMD (F04) in the corresponding bottom $panels$.} 
\label{hiacs}
\end{figure}

A more detailed insight on the metal-rich populations of $\omega$~Cen can be
achieved by observing the average metallicity of stars belonging to the
different branches in the $ACS~sample$ (see Fig. \ref{hiacs}). It is worthy of notice that 
stars selected along different branches in the CMD are confirmed to belong to different 
metallicity groups. In particular, stars standing on
the two brightest SGBs have a mean metallicity of [Fe/H]$\sim$~--1.7 and
[Fe/H]$\sim$~--1.4 respectively, in full agreement with the results obtained
using the $WFI~sample$. Stars located on the two faintest SGBs have an average 
metallicity of [Fe/H] $\sim$ -1.1 and [Fe/H] $\sim$ -0.6, respectively. By
comparing the obtained metallicities with the metallicity groups identified on
the RGB of $\omega$ Cen by S05, we can associate the dominant metal-poor
group ([Fe/H] $\sim$ -1.7) to the RGB-MP population and the secondary metal-rich
one ([Fe/H] $\sim$ -1.4) to the RGB-MInt2 population. Concerning the group of
stars with [Fe/H] $\sim$ -1.1, S05 found a bulk of RGB stars with similar
metallicity. A direct comparison with the existing high-resolution
spectroscopical analysis on RGB stars in $\omega$~Cen confirms the
presence of a number of stars with metallicity [Fe/H] $\sim$ -1.0 (Norris \& Da
Costa 1995, Vanture et al. 2002, P02, Pancino 2003). Given its metal content, we
can therefore associate this population of SGB stars to the RGB-MInt3
population, as defined by S05. SGB-a stars, located on the faintest SGB, have a
significantly higher metallicity ($\sim$ 1 dex) with respect to the bulk
population, representing the extreme metal-rich extension of the stellar content
of $\omega$ Cen, fully consistent with the measures obtained by P02 for RGB-a
stars. {\it This result represents the first direct confirmation that the RGB-a
and the SGB-a sequences belong to the same population}. For clarity, in the
following we will adopt for the different populations the naming convention
defined by S05. The adopted metallicities for the
observed populations of $\omega$ Cen are summarized in Table~2.

\section{Relative ages}

The SGB-TO is the most sensitive region of the CMD to the age of a stellar population.
Once the metal content of each SGB population has been measured, we should be
able to derive relative ages for each population through an appropriate
comparison with theoretical isochrones.

We used a set of theoretical isochrones calculated adopting the most up-to-date
input physics (Straniero, Chieffi \& Limongi 1997). In particular, the equation
of state includes the electrostatic correction (see Prada Moroni \& Straniero
2002) and microscopic diffusion (gravitational settling and thermal diffusion).
The theoretical isochrones have beed transformed into the observational planes 
by means of the synthesis code described in Origlia \& Leitherer (2000) and 
using the model atmospheres by Bessel, Castelli \& Plez (1998, hereafter BCP98).
For the ACS data set, the filter responses and camera throughputs, kindly 
provided by the ACS User Support Team, have been used.

To compare the observed SGBs with theoretical isochrones, we need to assume a
distance modulus and a reddening correction. In the following we adopt the
distance modulus by Bellazzini et al. (2004), $(m-M)_0 = 13.70 \pm 0.13$. 
Concerning the reddening and extinction coefficients, we assumed $E(B-V)=0.11\pm
0.01$ (Lub 2002), $A_B=4.1~E(B-V)$, $A_R=2.35~E(B-V)$ (Savage \& Mathis 1979)
and $A_I=1.8~E(B-V)$ (Dean, Warren \& Cousins 1978).

\subsection{Chemical assumptions}

Although the evolution of a star along the SGB depends mainly on age
and metallicity, other factors may have a non negligible impact, namely the
$\alpha$-elements and helium abundance. Straniero \& Chieffi (1991) and Salaris, Chieffi \& Straniero (1993) showed
that, when computing the isochrones of Population II stars, the contribution of
the $\alpha$-element enhancement can be taken into account by simply rescaling
standard models to the global metallicity [M/H], according to the following
relation 

$$ [M/H] = [Fe/H] + log (0.638~\times~10^{[\alpha/Fe]} + 0.362) $$

The corresponding metallicity, in terms of mass fraction Z, is

$$ Z = (1-Y) \frac{10^{[M/H]+log[Z/X]_\odot}}{1+10^{[M/H]+log[Z/X]_\odot}} $$

where $[Z/X]_\odot = 0.0176$, according to Allende Prieto et al. (2002) and
Asplund et al. (2004). We adopted $[\alpha/Fe] = +0.3$ for both the MP and the
MInts population samples, according to the most recent high-resolution
spectroscopic results (Norris \& Da Costa 1995, Smith et al. 2000, Vanture et
al. 2002, P02). For the SGB-a stars, we adopted a significantly lower
enhancement $[\alpha/Fe] = +0.1$ as suggested by high-resolution optical spectra
(P02) and low-resolution IR spectra (Origlia et al. 2003). 

Regarding the helium abundance, several authors considered the effects of
differences in helium abundance among the different populations of $\omega$
Cen (F04, Bedin et al. 2004, D'Antona \& Caloi 2004, N04, P05, Lee et al.
2005). In particular, N04 ad Lee et al. (2005) suggested a helium enhancement
of $\Delta Y \sim 0.12$ for the MInt population and  $\Delta Y \sim 0.15$ for
the most metal-rich anomalous one, in order to reproduce the complex MS morphology 
of $\omega$ Cen. 
Although such high helium enhancements would imply a series of problems, 
concerning the helium yields for low-mass stellar systems (see \S 6), we explored this
possibility for the MInts and SGB-a populations by fitting the observed SGBs with
models having canonical He abundance and various He-enhancement levels. The
adopted metallicity, $[\alpha / Fe]$, Y and the derived age for each population
of $\omega$ Cen are summarized in Table~2.

\subsection{Results}

\begin{figure}
%\plotone{f5.eps}
\plotone{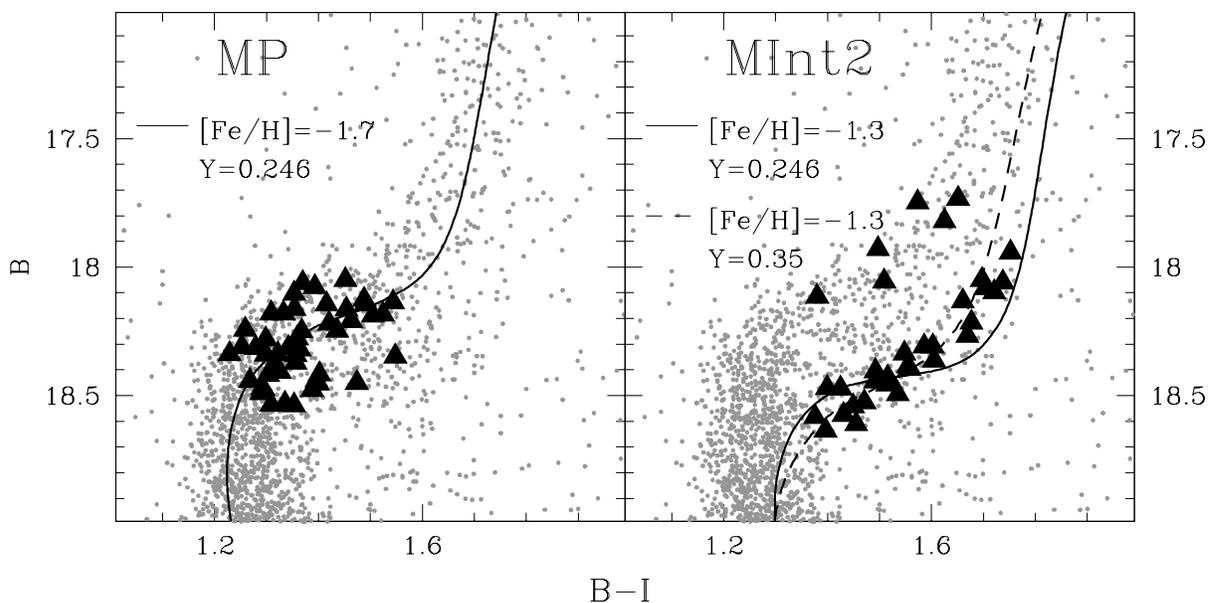}
\caption{Isochrone fitting of the SGB sub-populations in the
$WFI~sample$. Stars with 
metallicity $[Fe/H]<-1.8$  ($SGB-MP$) and $-1.4<[Fe/H]<-1.0$ 
($SGB-MInt2$) are plotted as triangles in the {\it left} and 
{\it right panel}, respectively. 
Theoretical 16 Gyr old isochrones with appropriate metallicity and helium abundance
(Y=0.246: {\it solid lines} and Y=0.35: {\it dashed line}) are overplotted.} 
\label{to}
\end{figure}

\begin{figure}
%\plotone{f6.eps}
\plotone{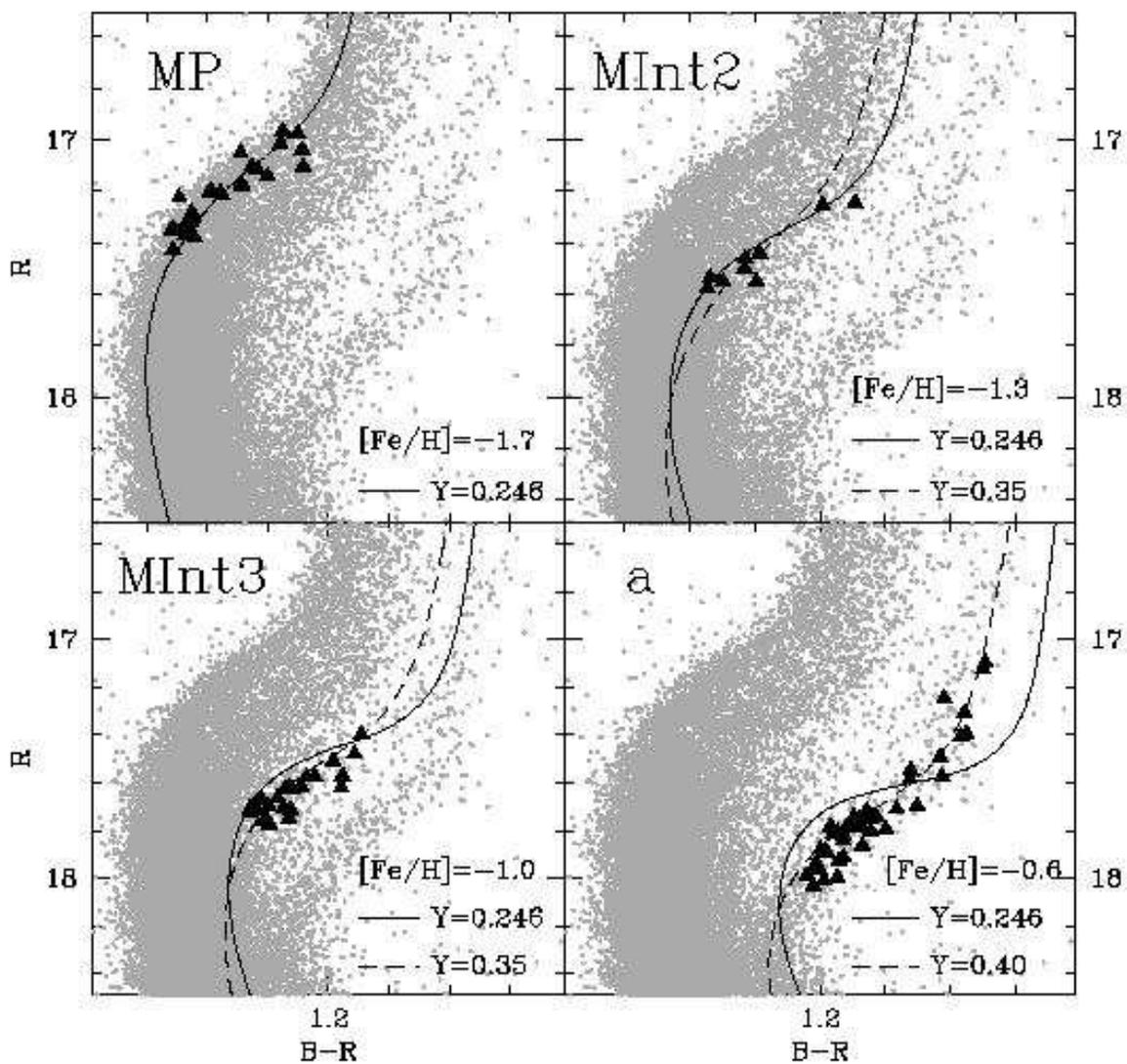}
\caption{Isochrone fitting of the four observed SGBs of $\omega$ Cen in the
$ACS~sample$. The spectroscopic target stars belonging to different
sub-populations are marked on the CMD as triangles.
Theoretical 16 Gyr old isochrones with appropriate metallicity 
and helium abundance (Y=0.246:solid lines and different enhanced helium
abundances: dashed
lines) are overplotted.} 
\label{toacs}
\end{figure}

\begin{figure}
%\plotone{f7.eps}
\plotone{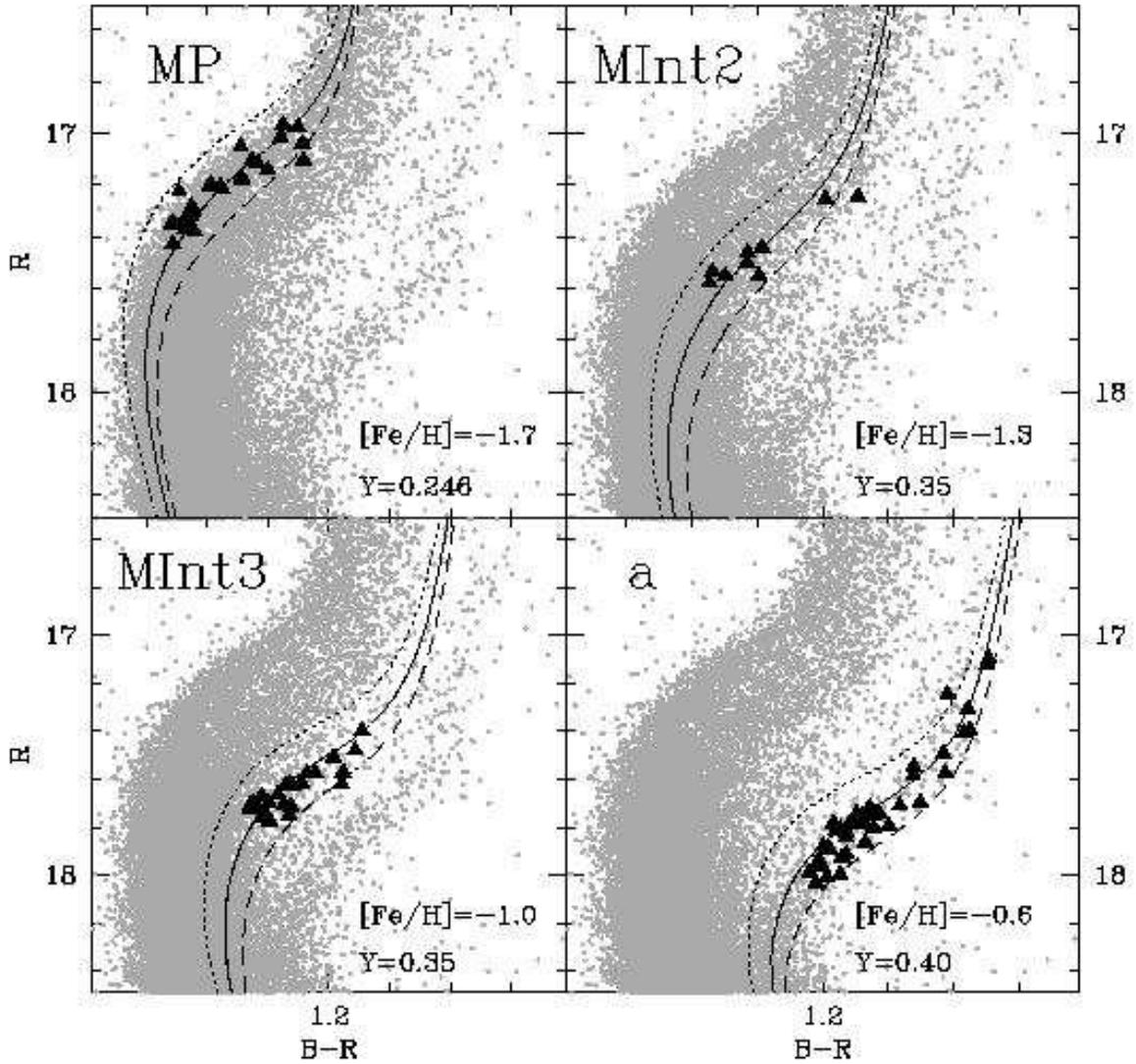}
\caption{Same as Figure 6.
Theoretical isochrones with appropriate metallicity and helium abundance (see
Table 2) and various ages (14 Gyr dotted lines; 16 Gyr solid lines; 18 Gyr
dashed lines) are overplotted.} 
\label{iso}
\end{figure}

Fig. \ref{to} and \ref{toacs} show the isochrone fitting for the observed
populations of $\omega$ Cen for the $WFI~sample$ and the $ACS~sample$
respectively, assuming for each population different helium abundances ranging
from the cosmological value Y=0.246 (Salaris et al. 2004; Cyburt et al. 2003) up
to Y=0.40 . Regarding the $WFI~sample$, in
order to perform a meaningful comparison and to avoid spurious contamination
from stars with uncertain metallicity measurement, we considered only stars with
[Fe/H]$<-1.8$  (SGB-MP population) and $-1.4<$[Fe/H]$<-1.0$ (SGB-MInt2
population).
As can be seen from Fig. \ref{to}, apart from the presence of few outliers
(probably due to spectroscopic and/or photometric errors), there is a clear
segregation between the two samples. While most of the SGB-MP stars 
populate the brighter portion of the SGB, SGB-MInt2 stars appear to be mainly 
located in the fainter part of the SGB. The two groups of stars are fit by
theoretical isochrones of the same age (16 Gyr) and appropriate metallicity (see
\S 4.1). In Fig. \ref{to}b an isochrone with a significant helium
enhancement (Y=0.35) is also overplotted. As can be seen, the location of the
target stars is nicely reproduced by this isochrone.
The same procedure has been followed for the $ACS~sample$ (see Fig. \ref{toacs}). 
Also in this case a set of 16 Gyr old
isochrones with appropriate metallicities and different helium enhancements are 
overplotted. As can be noted, the adoption of a 
different helium abundance yields a better fit 
of the SGB shape but does not significantly affect the relative 
ages. Note that all the populations are nicely reproduced by isochrones with the 
same age. In order to further support this result, we plotted the four isochrones
with the best choice of helium abundance, as derived from the comparison shown
in Fig. \ref{to}b and \ref{toacs}, and various ages in Fig.~\ref{iso}. As can be
seen, a change of $\pm 2 Gyr$ in the adopted age produces a significant
variation in the isochrone location. Hence, the average error in the age
differences, estimated on the basis of the effects 
produced on the estimated age by the intrinsic magnitude spread of the observed 
SGBs, can be assumed to be $\sim 1.5~Gyr$.
Summarizing, we reproduced all the 
four considered populations with 16~Gyr old isochrones. The
adopted He abundances are: the cosmological Y=0.246 for the SGB-MP population,
Y=0.35 for the SGB-MInt2 population, Y=0.35 for the SGB-MInt3 population and Y=0.40
for the SGB-a population. 
We found that a large helium enhancement affects the morphology of the SGB but
does not change its average magnitude. In particular, increasing the helium
abundance:

\begin{itemize}

\item The slope of the SGB becomes steeper;

\item The color difference between the TO and the RGB base decreases.

\end{itemize}

{\it As a consequence, the assumption of a differential helium
abundance among the different populations of $\omega$ Cen allows a better fit 
of the observed SGB morphology but does not significantly affect the relative 
ages}.

Note that although the old zero point of the adopted age scale differs from the 
most recent determinations of the age of the universe (Cyburt et al. 2003), 
the age differences are marginally affected by this assumption.

\begin{figure}
%\plotone{f8.eps}
\plotone{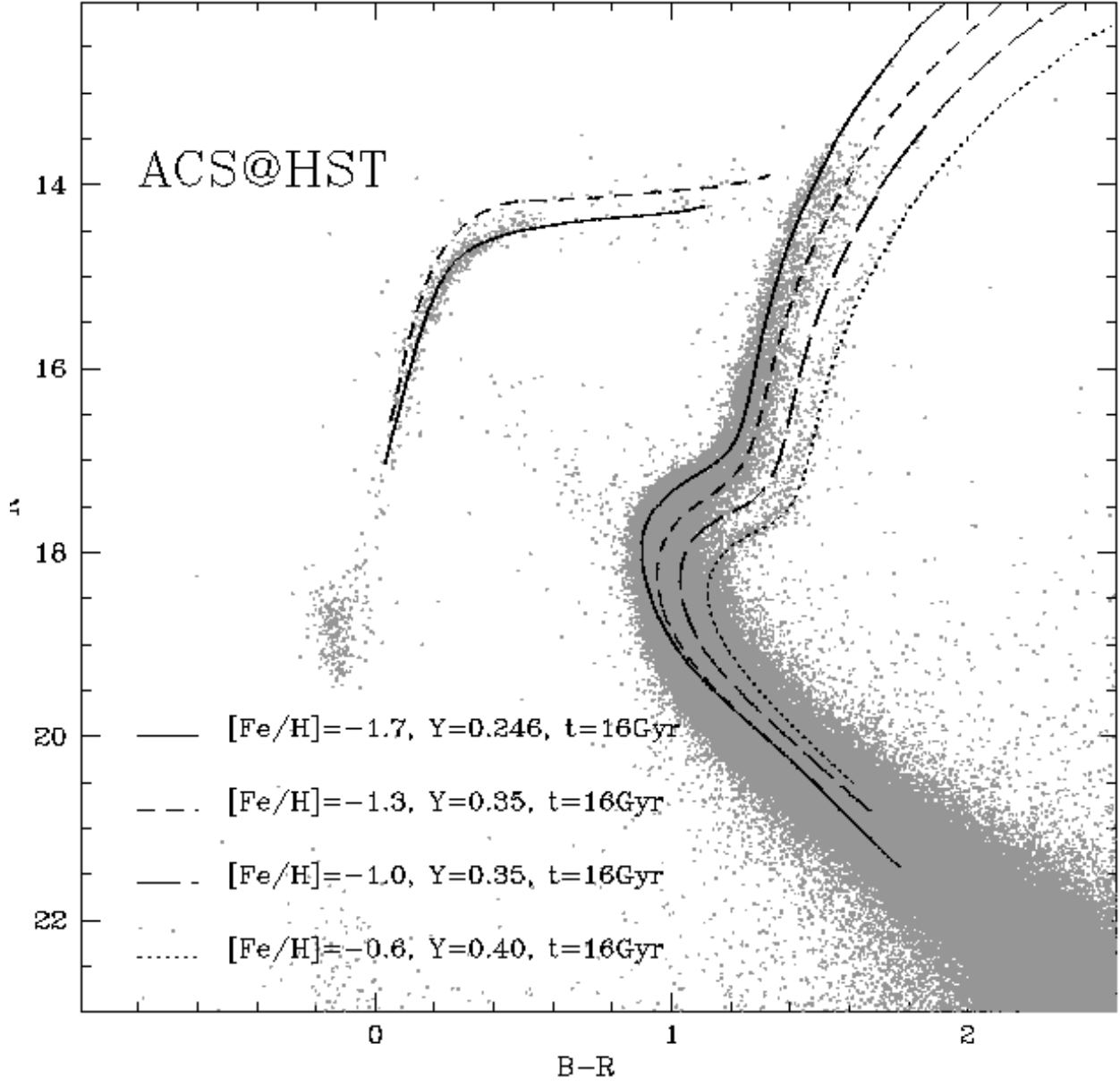}
\caption{ACS CMD of $\omega$ Cen. Theoretical isochrones with the bestfit
parameters indicated in \S 4.2 are overplotted.} 
\label{cmd}
\end{figure}

In Fig. \ref{cmd} the four isochrones are overplotted to the entire ACS CMD of
$\omega$ Cen. The isochrones correctly reproduce the global shape of the 
observed sequences from the MS up to the RGB. 
However, is worth of noticing that:  

{\it (i)} Our best-fit isochrone for the MInt2 population, having Y=0.35, does not cross the MP isochrone at the MS level. If we
accept the hypotesis that the MInt2 population is related to the blue MS
observed by Bedin et al. (2004), we need to adopt a larger helium
abundance ($Y_{MInt2} > 0.35$) in agreement with what found by N04, P05
and Lee et al. (2005). 

{\it (ii)} Together with isochrones, the Zero Age Horizontal Branch (ZAHB) of the MP and
MInt2 populations are plotted on the CMD. We used the latest models computed
using FRANEC (see Straniero, Chieffi \& Limongi 1997 for details) \footnote{The 
ZAHB were calculated by interpolating the location in the CMD of the tracks of stars with 
masses ranging from 0.52 to 0.80 $M_\odot$. The core masses and the surface
compositions were derived from the corresponding last models in the H-burning shell. 
The ZAHB model was set when all the secondary elements in 
their H-burning shell are relaxed to their equilibrium values.}. As can be immediately noted, the ZAHB
corresponding to the MInt2 population is significantly brighter than the
observed HB. This is a consequence of the large helium abundance assumed for
this population, that strongly increases the CNO cycle efficiency in the HB
stars envelopes and leads to a stable equilibrium stage at higher luminosity.
Butler et al. (1978) and Rey et al. (2000) found a group of RR Lyrae variables
with metallicity significantly higher than that of the dominant cluster
population located $\sim 0.3~mag$ below the average RR Lyrae V magnitude.
If we associate the metal-rich RR Lyrae to the MInt2 population, the assumption
of a significant increase in the He abundance would produce a stark incongruence
with the observed luminosity of such metal-rich RR Lyrae.
Similar incongruences between the expected and the observed
 luminosity of metal-rich RR Lyrae stars have been noted by N04 and Sollima et al.
(2005b in preparation). 
In this respect, note that most of the helium-rich HB stars reach
higher temperatures because of their smaller masses, and therefore populate
essentially the blue tail of the HB, a region where magnitude differences cannot
be appreciated (P05, Lee et al. 2005). Hence, only few helium-rich stars are expected to cross the instability
strip.   

\section{Kinematics}

\begin{figure}
%\plotone{f9.eps}
\plotone{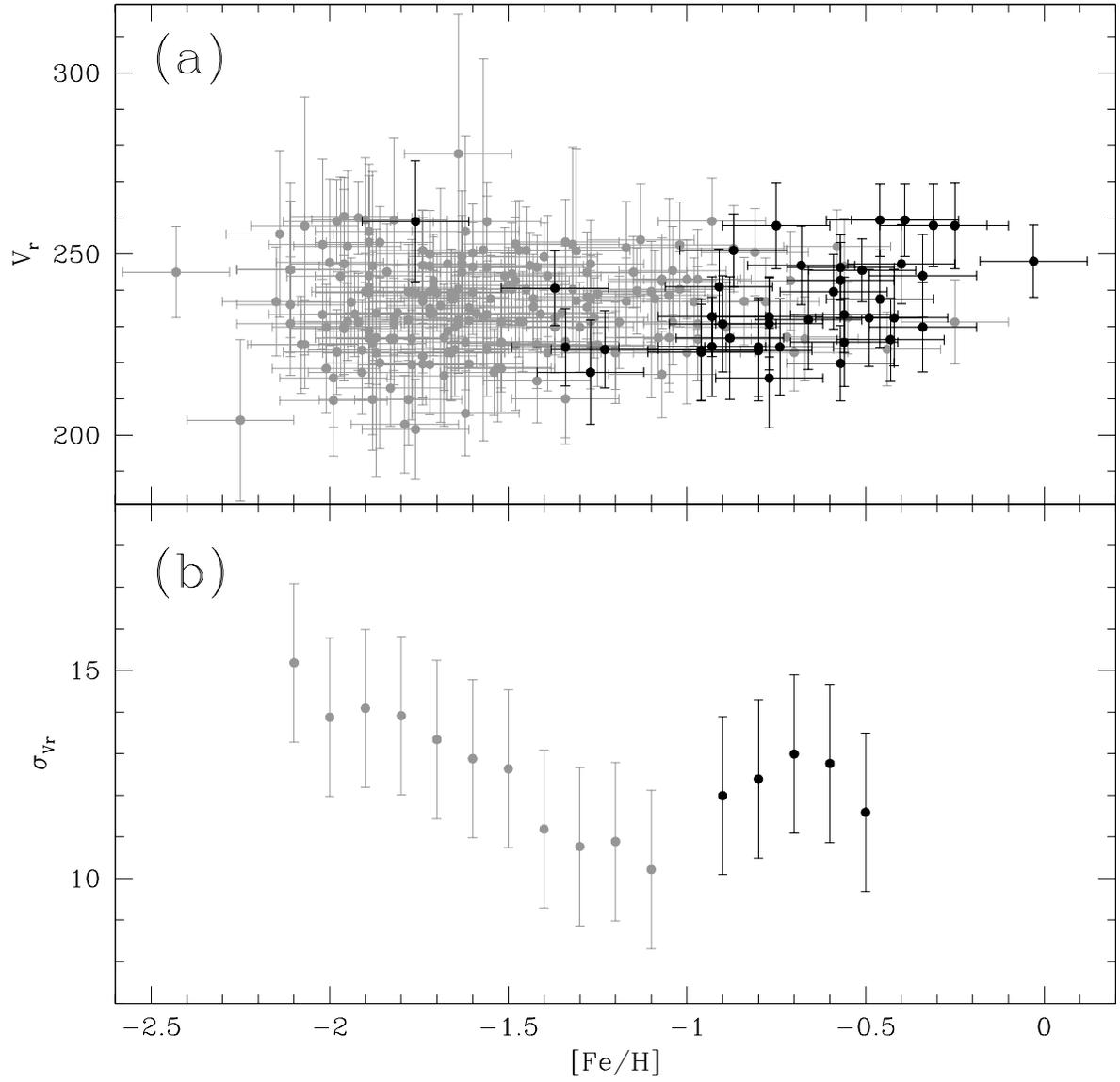}
\caption{Radial velocity as a function of metallicity ($panel$ a). The 
velocity dispersion is plotted against metallicity in $panel$ b. SGB-a
stars are marked in both panels with black symbols.} 
\label{vel}
\end{figure}

Another intriguing characteristic of $\omega$ Cen is the observed connection
between kinematics and abundances. From Ca II triplet analysis of about 400 RGB
stars, 
Norris et al. (1997, hereafter N97) found that the 20\% metal-rich tail
of the metallicity distribution is kinematically cooler than the 80\% metal poor 
component. Beside this result, N97
found that their metal-rich component does not show the same systemic rotation
as the metal-poor one. This last finding appears in contradiction to a simple
dissipative scenario. We used our sample to find further evidences of such
behaviour. Fig. \ref{vel} shows the radial velocity and velocity
dispersion as a function of metallicity for the global sample of stars presented
in this paper. The mean radial velocity of the whole sample is
$v_r=-235.9\pm0.8~Km~s^{-1}$ and the average dispersion
$\sigma_{v_r}=12.6~Km~s^{-1}$. SGB-a stars
present no significant radial velocity offset with respect to the bulk
population stars, having $v_r^{SGB-a}=-236.5\pm1.7~Km~s^{-1}$.

A surprising result concerns the behaviour of the velocity dispersion as a function
of metallicity. As can be seen from
Fig.~\ref{vel}b, for stars with $-1.7~<[Fe/H]<~-1.0$, the velocity dispersion
decreases as the metal abundance increases, in agreement with the result by N97. 
SGB-a stars do not follow this
trend: they show a significantly larger velocity dispersion and seem to be
kinematically warmer than the MInt stars. This evidence is confirmed even if we
consider only stars lying in the inner $5'$, and seems to be
independent on the distance to the cluster center. Although the sample used by N97 spans a
metallicity range similar to that covered by the present analysis (-1.6
$<~[Ca/H]~<$ 0.0), it contains only a few stars at [Ca/H] $>$ -0.7. For this
reason, while at metallicity $-2.0 <~[Fe/H]~< -1.0$ the trend observed by N97 is
fully confirmed by our analysis, the peculiar kinematical behaviour of SGB-a
stars could not be detected by N97. Summarizing, the MInt populations are the
only cool components, being kinematically cooler than both the (hot) MP
population and the (warm) SGB-a. The observed behaviour of the kinematical
properties of $\omega$~Cen as a function of its metallicity is still far from
being fully understood.  

The warm velocity dispersion of the SGB-a stars adds up to other
structural and kinematical peculiarities of the RGB-a stars. First, the RGB-a
centroid in the spatial density distribution appears significantly dislocated
from the main population's one (Pancino et al. 2003, S05). Moreover, the mean
proper motion of RGB-a stars differs significantly from that of the main
cluster population (Ferraro et al. 2002). This last result has been questioned
by Platais et al. (2003) who claimed that the different proper motion observed
for RGB-a stars is due to a spurious instrumental
effect in the original proper motion catalog by van Leuween et al. (2000).
However, while the arguments of Platais et al. (2003) were already considered
and dismissed in the original publication by van Leuween et al. (2000), they
also have been strongly criticized by Hughes et al. (2004), who quantitatively
demonstrated that the original proper motion catalogue is free from any spurious
instrumental trend. Further support
to the correctness of the proper motion measurements has been brought by Pancino
(2003), who again did not found any significant proper motion variation with
neither magnitude nor color.

\section{Discussion and Conclusions}

We presented low-resolution spectroscopy of 256 SGB stars in $\omega$ Cen.
Metallicities and radial velocities have been measured from Ca II triplet lines
analysis. The metallicity distribution function confirms the metallicity spread
observed in previous analysis performed on giant stars. SGB-a stars appear to
have a metallicity higher than the bulk population, representing the extreme
metal-rich extension of the stellar content of $\omega$ Cen. The relative ages
of the different stellar populations of the system have been estimated by
fitting the observed SGBs with theoretical isochrones having appropriate
metallicities and various helium content. The ages derived for the different
populations are all compatible within 2 Gyr. This result indicates that
$\omega$ Cen enriched itself in a short timescale ($< 2~Gyr$) and imposes firm
constraints on the chemical evolution of the system.

The hypotesis of a large helium gradient between the populations of $\omega$ Cen
has been tested. We found that the morphology of the SGB of the most metal rich
populations of the cluster can be better reproduced assuming a helium 
enhancement, with respect to the dominant metal poor population, of $\Delta Y
\sim 0.10-0.15$ for the MInts population and $\Delta Y \sim 0.15-0.20$ for the
anomalous one, in good agreement with previous estimates in literature. Such
large helium overabundances do not affect significantly the relative
ages of the stellar populations.

The results presented here exclude an extended star formation
period, regardless of the adopted helium
abundances. The average error in the age differences is $\sigma_{\Delta t} \sim
1.5~Gyr$ (see \S 4.2), so age differences smaller than this quantity cannot be appreciated. 
Nevertheless, this finding imposes serious problems related
to the chemical enrichment history of this stellar system. Hilker \& Richtler
(2000) claimed an extended star formation period in order to explain the tight
correlation between CN-band strengths and iron abundances. Moreover, the
abundance patterns of s-process elements require the contribution of low-mass
AGB stars ($1.5 \div 3 M_{\odot}$, Smith et al. 2000). Stars of these masses
reach the AGB phase after 0.7 - 1.5 Gyr that should represent a lower limit for the
age difference between the populations of $\omega$ Cen. Finally, in the
hypotesis of a large helium enhancement, the derived helium abundance gradient
between the MP and MInt2 population turns out to be $\delta Y/\delta Z \sim
180$, in stark contrast with more canonical values of  $\delta Y/\delta Z \sim
3-4$ (Pagel et al. 1992, Jimenez et al. 2003). An efficient He production driven 
by low-mass AGB stars would be accomplished by a corresponding C overabundance, 
in contradiction with observational evidences (Smith et al. 2000). In the same
way, intermediate mass stars would produce, beside He, an unusual N abundance. 
A larger helium yield is provided by more massive objects (M $>~20 M_{\odot}$) 
wherein black hole production limits the amount of heavy elements relative to 
helium in the ejecta released into the interstellar medium (Timmes et al. 1995). 

The kinematics of the stellar populations of the cluster appear more puzzling
than ever. The velocity dispersion as a function of metallicity shows a mimum at
$[Fe/H] ~\sim~ -1.1$ ,
suggesting that MInt populations are kinematically cooler than both the MP and the
anomalous population.

The emerging scenario can be summarized as follows:

i) The MP population ([Fe/H] $\sim$ -1.7) represents the first and largest event
of star formation of $\omega$ Cen. It contains the majority of cluster's stars,
and can be well reproduced by an old isochrone having a cosmological helium
abundance. Moreover, it presents the highest velocity dispersion among the
different cluster populations.

ii) The MInt2 population ([Fe/H] $\sim$ -1.3) could have had a peculiar history.
It is suspected to be connected to the bMS (Bedin et al. 2004, P05), and for
this reason it requires an anomalous helium overabundance (Y $>$ 0.35). This
population formed after a short timescale ($<$ 2 Gyr) from a medium enriched by
SNII in iron and $\alpha$-elements abundance, and by AGB winds of massive stars
in s-process elements and, possibly, helium (see also the discussion in \S 4.2). 
How a such huge amount of helium
could have been produced remains at present an unanswered question. Moreover,
MInt2 stars present a smaller velocity dispersion with respect to the bulk metal
poor population ones, indicating an evolution in the dynamical settling of the
system.
The MInt3 population ([Fe/H] $\sim$ -1.0), together with MInt2,
represents the kinematically cool component.

iii) The anomalous population represents the metal-rich end of the stellar
population mix of $\omega$ Cen at [Fe/H] $\sim$ -0.6. Although increasing
the Helium abundance to Y$\sim$0.4 improves the fit of the SGB-a shape, no
value of Helium abundance can make this anomalous population significantly
younger than the MP population. Moreover, the analysis of radial velocities of
these stars indicates a kinematical behaviour different from the MInt
populations. For these reasons, the understanding of the origin and evolution of
this population is still a mistery. 

Two different scenarios can be put forward
to explain its origin:

a) A very fast self-enrichment process generated the anomalous population in the
early stage of evolution of $\omega$ Cen, producing a large amount of helium
($\Delta Y \simeq 0.4$), s-process elements ([s/Fe] $\simeq$ +1.0) and iron
(leading to a lower [$\alpha$/Fe]$ \simeq$ +0.1) in less than 2 Gyr.

b) The anomalous population evolved in a different environment and was
later accreted by the main body of $\omega$~Cen.

Freyhammer et al. (2005) suggested that
stars belonging to the SGB-a are physically separated from and more distant than
the main body of $\omega$~Cen by about 120-250 pc. To achieve their best fit
under this hypothesis, however, the authors have to assume for the SGB-a a
metallicity of --1.1$<$[Fe/H]$<$--0.8, in stark contrast with the present
determination, and the abundance measured along the RGB-a ([Fe/H]$\simeq$--0.6$\pm$0.15, P02 and
Origlia et al. 2003). Moreover, the few radial velocity determinations of RGB-a
stars (P02, Vanture et al. 2002 and Origlia et al. 2003) and the present large
sample of SGB-a measurements suggest that the radial velocity of this population
is not different from the $\omega$~Cen systemic velocity, therefore
implying that SGB-a stars and $\omega$~Cen are at least dynamically bound to each
other.

The results presented in this paper confirm once more that a simple
self-enrichment scenario cannot explain all the observational evidences gathered
in the past. Many questions remain to be understood in order to explain the
formation and evolution of this peculiar stellar system. Therefore, it is clear
that if one desires to accommodate all the observational facts in one
self-consistent scenario, it is necessary to include not only a complex chemical
evolution, but also a complex dynamical history.

\acknowledgments

This research was supported by the Ministero dell'Istruzione, dell'Universit\`a
e della Ricerca. We warmly thank Paolo Montegriffo for assistance during the
catalogs cross-correlation and astrometric calibration process and Luciano
Piersanti for his helpful support in the theoretical tracks calculations.
We thank also the anonymous referee for his helpful comments and suggestions.
AS acknowledge the {\it Marco Polo Project} for the financial support and
the European Southern Observatory in Garching for the hospitality during his stay.

\appendix

\section{ Calibration of the $\Sigma~Ca$-[Fe/H] relation}

In this appendix we present the $\Sigma~Ca$ index calibration as a function of
metallicity for sub-giant stars, as derived from the application of the {\it
running box technique} to the most extensive catalog of spectra available in 
literature (C01). 

\begin{figure}
%\plotone{f10.eps}
\plotone{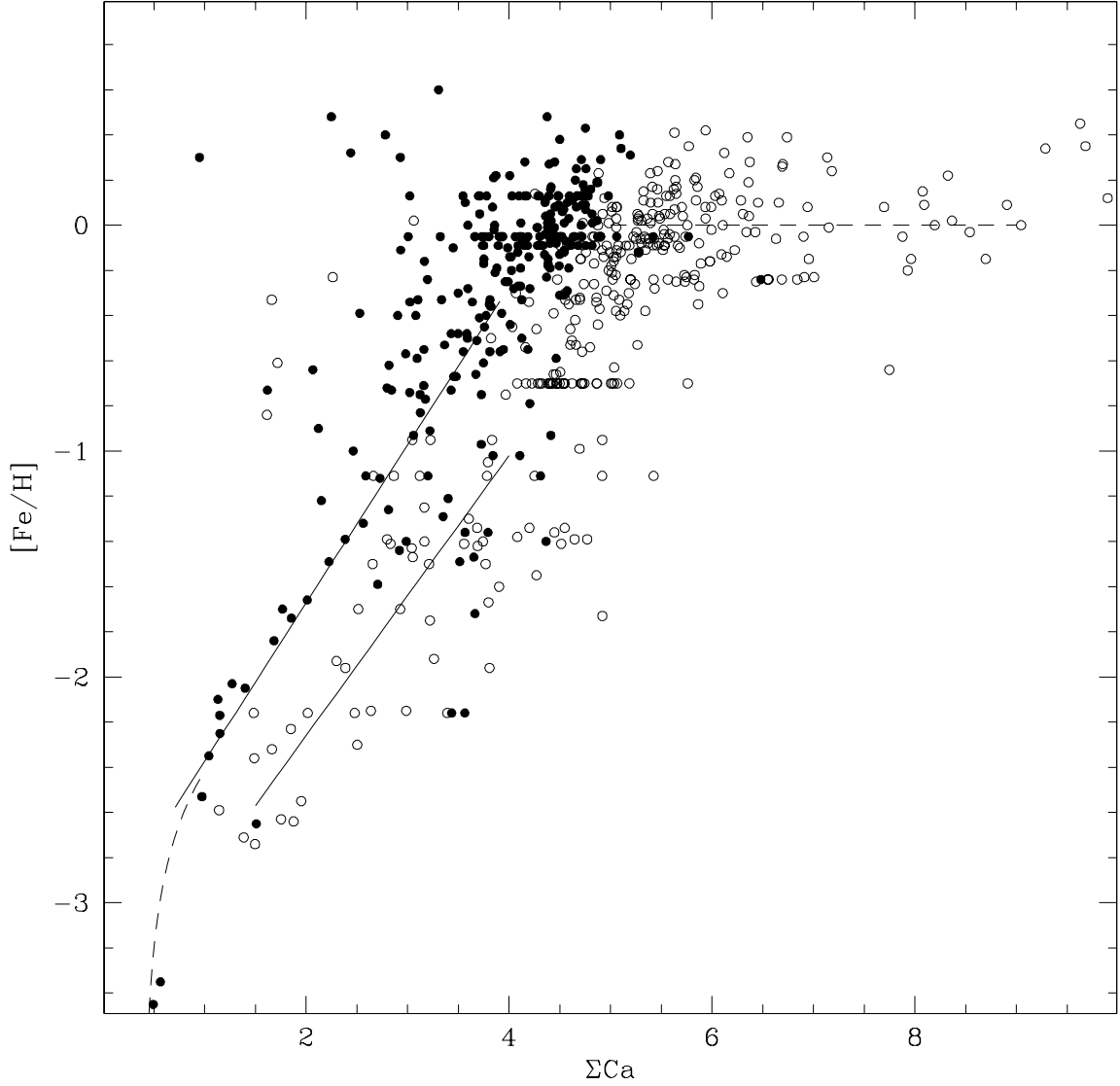}
\caption{Metallicity dependence of the $\Sigma~Ca$ index for the 603 stars of
C01. Open points indicate stars with $log~g < 3$, filled points indicate stars
with $log~g > 3$. The linear behaviour of the 
$\Sigma~Ca$ index as a function of metallicity is indicated with solid lines.
Dashed lines represent the behaviour of the $\Sigma~Ca$ index in the range where it does not
correlate linearly with metallicity.} 
\label{cen}
\end{figure}

Since the EW measurement technique and index definition by C01 are slightly
different from the {\it equivalent line strength index} $\Sigma~Ca$ defined in
\S 2.2, as first step we measured $\Sigma~Ca$ as defined in \S 2.2 on the 603
spectra of C01. Fig.~\ref{cen} shows the location of the C01 stars in the
$\Sigma~Ca$ -- [Fe/H] plane, where different symbols are used for stars in
different gravity ranges. As expected, in the metallicity range
--2.8$<$[Fe/H]$<$-0.4, the $\Sigma~Ca$ index increases approximately linearly
with metallicity. For stars with higher metallicity ([Fe/H] $>~-0.3$) the
$\Sigma~Ca$ index depends mostly on gravity. For very metal-poor stars
([Fe/H] $<~-3$) the Ca II triplet lines strength become insignificant, making it
difficult to measure in noisy spectra. In this metallicity range $\Sigma~Ca$ is
expected to correlate with metallicity in an asymptotic way. 

The $\Sigma~Ca$ index measured in the $\omega$~Cen targets cover the range 
0.8\AA$<\Sigma~Ca<$3.7\AA. In this range the dependence of metallicity on
$\Sigma~Ca$ can be reasonably approximated by a linear relation. However, the effect
of gravity is non-negligible and can significantly affect such a relation. For
this reason, we decided to further investigate the parameters space occupied by
our target stars in the ($T_{eff}, \log~g$) plane.

\begin{figure}
%\plotone{f11.eps}
\plotone{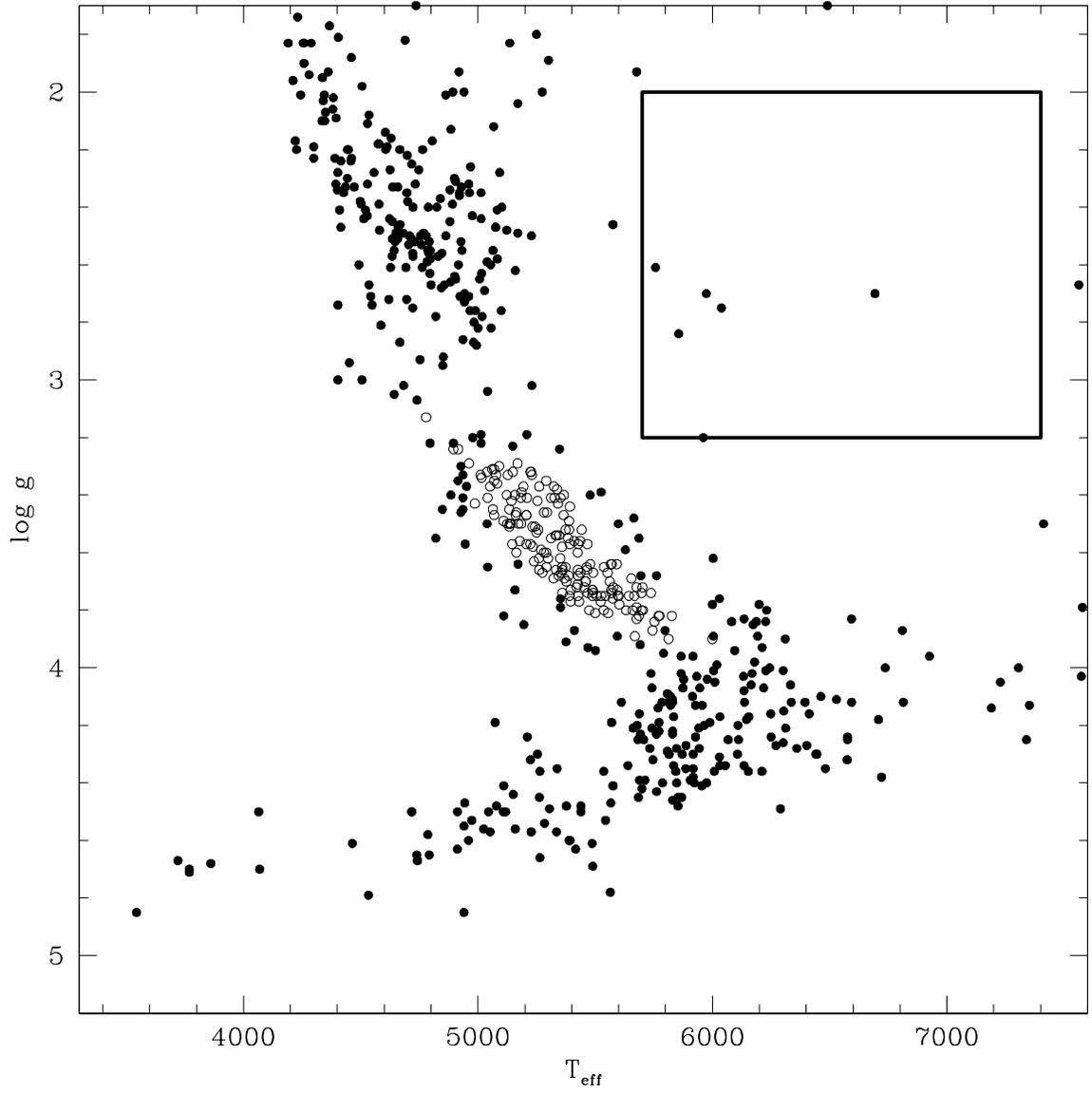}
\caption{$log~g - T_{eff}$ plane. The calibration stars by C01 (filled circles) 
and $\omega$ Cen target stars (open circles) are shown. 
The size of the selection box used for the $\Sigma~Ca - [Fe/H]$ 
calibration is indicated.} 
\label{box}
\end{figure}

Temperatures were derived using the color-temperature conversion provided by 
BCP98. Gravities were derived from the universal gravitation law assuming a mass
of 0.8 $M_{\odot}$ (Bergbusch \& VandenBerg 2001) and the temperatures derived
above. We used the BVI magnitudes by P00 for the $WFI~sample$, while for the 
$ACS~sample$ stars we used the BR magnitudes by F04 in order to derive 
temperatures and gravities, once the BCP98 conversions have been corrected for
the ACS filter response and camera troughputs (see \S 4). Fig. \ref{box} shows
the location of the target stars of $\omega$~Cen ({\it open circles}) with
respect to the C01 stars ({\it filled circles}) \footnote{Before applying the calibration, 
all spectra were smoothed to the resolution of the calibration spectra of C01 
($FWHM \sim 1.5$ \AA) and the analysis was repeated in order to measure the 
shift in the EW measurement due to the different resolution. 
We find a small $\Sigma~Ca$ shift of $\Delta \Sigma~Ca = 0.044 \pm 0.062$ \AA, 
that was applyied to all the $\Sigma~Ca$ measurements and its error was 
propagated into the final $\Sigma~Ca$ error.}.
As can be seen, the bulk of the
target stars covers the range $3 < \log~g < 4$ and $4600~K < T_{eff} < 6000~K$.
Only a bunch of C01 stars are present in this range. For this reason we decided
to apply the so-called {\it running box technique} to the ($T_{eff}, \log~g$)
plane in order to derive appropriate $\Sigma~Ca$ -- [Fe/H] relations. A sampling
box with $\Delta T_{eff}=1700~K$ and $\Delta \log~g=1.2~dex$ has been defined
and moved in steps of $\delta T_{eff}=400~K$ and $\delta log~g=0.3~dex$,
scanning the entire area covered by the C01 sample in the ($T_{eff},
\log~g$)-plane (Fig.~A2). All the C01 stars
(with EW in the range 0.8 \AA $<\Sigma~Ca<$3.7 \AA) lying in each box have been
used to derive a linear regression relation between $\Sigma~Ca$ and [Fe/H].
Finally, a general fitting function for the whole parameter space has been
constructed by smoothing the derived local functions with a gaussian filter. For
a given point in the parameter space with coordinates $P(\log~g, T_{eff},
\Sigma~Ca)$ we computed the corresponding value of metallicity $[Fe/H]_{i}$, for
each box containing it. A gaussian-weighted average of the various $[Fe/H]_{i}$
corresponding to all the boxes containing $P$ yields the final metallicity,
defined as follows:

$$ [Fe/H] = \frac{\sum_{i} w_{i}~[Fe/H]_i}{\sum w_{i}}$$

where the weight $w_{i}$ is modulated by the distance of the considered $P$ from 
the center of the i-th box in the ($log~g, T_{eff}$)-plane

$$w = e^{-(\frac{log~g-log~g_{i}}{\sigma_{g}})^2-(\frac{T-T_{i}}{\sigma_{T}})^2}$$
$$\sigma_{g}=1.2~dex$$
$$\sigma_{T}=1700~K$$ 

\begin{figure}
%\plotone{f12.eps}
\plotone{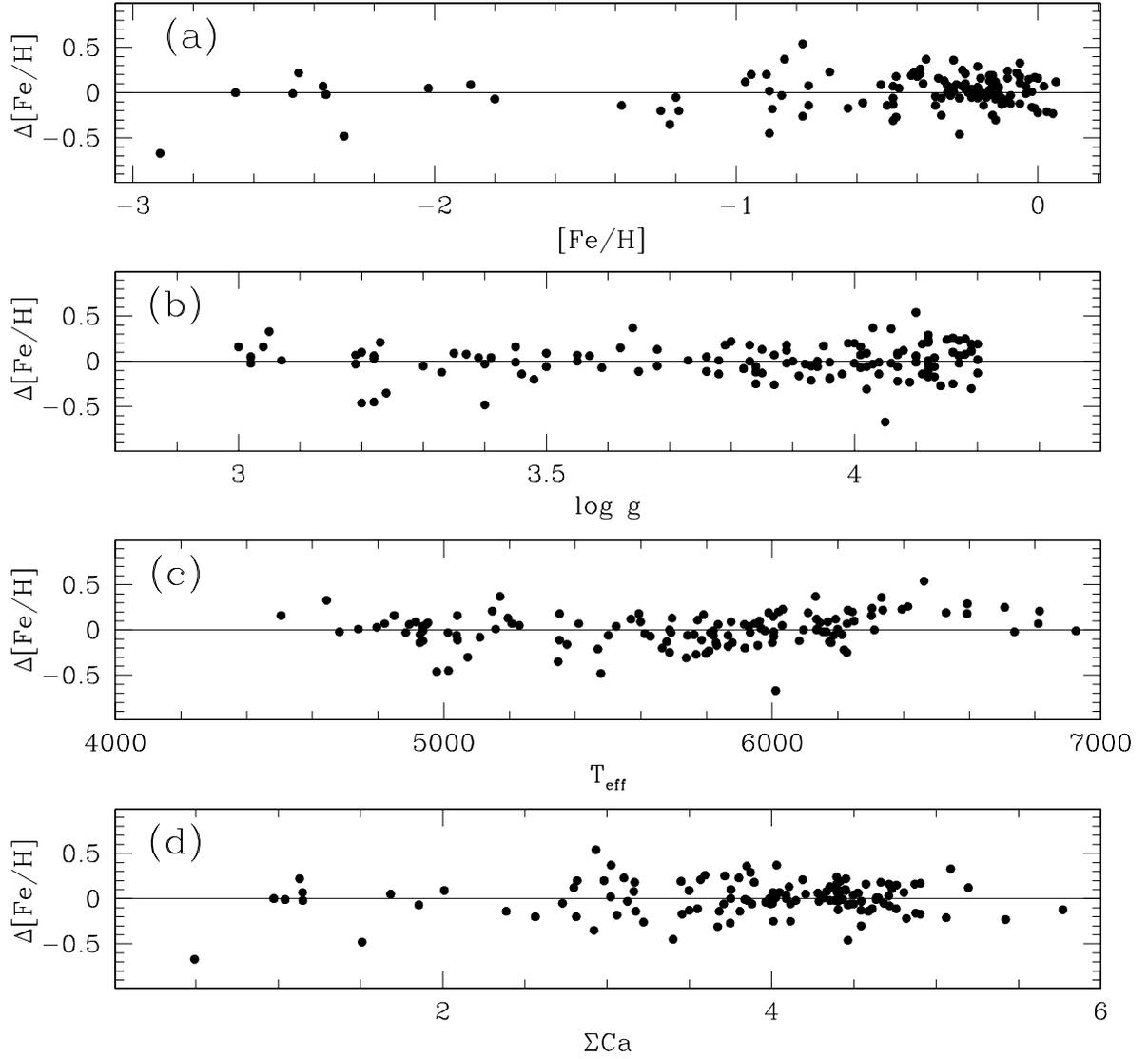}
\caption{Residuals of the metallicity derived following the procedure described in this
appendix with respect to the values
listed by C01 as a function of [Fe/H] ($panel$ a), log g 
($panel$ b), $T_{eff}$ ($panel$ c) and $\Sigma~Ca$ ($panel$ d).} 
\label{res}
\end{figure}

In order to check the reliability of the present calibration, we derived the
metallicity of all the C01 sample stars using the procedure described above.
Fig.~\ref{res} shows the difference of metallicity between our determination and
the original determination by C01, as a function of the involved parameters
($[Fe/H]$, $\log~g$, $T_{eff}$ and $\Sigma~Ca$, respectively). As can be seen,
no significant residual trends are present. The mean dispersion of the fit is
0.15 dex, that can be assumed as the overall accuracy of the calibration.

\subsection{Metallicity errors}

Several sources of uncertainty occur in the present metallicity determination. 
In particular:

\begin{itemize}

\item Uncertainties in the EW measurements;

\item Uncertainties in the resolution correction;

\item Photometric errors on the BVRI magnitudes;

\item Uncertainties in the color-temperature and\\ 
      magnitude-gravity conversion;

\item Uncertainties in the $\Sigma~Ca$-[Fe/H] calibration.

\end{itemize}

In order to estimate the overall uncertainty in the derived metallicities, we
performed a Monte Carlo simulation spanning a range of $\pm 5 \sigma$ in each
error source listed above. We assumed the uncertainties on the $\Sigma~Ca$
as described in the \S 2.2, the photometric errors given by P00 and F04, the
temperature uncertainty reported by BCP98 and propagated into the gravity
determination, and the error in the $\Sigma~Ca$-[Fe/H] calibration as derived in
the previous section. The resulting mean error is $\sigma_{[Fe/H]}=0.2~dex$.

\clearpage

\clearpage

\clearpage

\begin{figure}
%\plotone{f3.eps}
\plotone{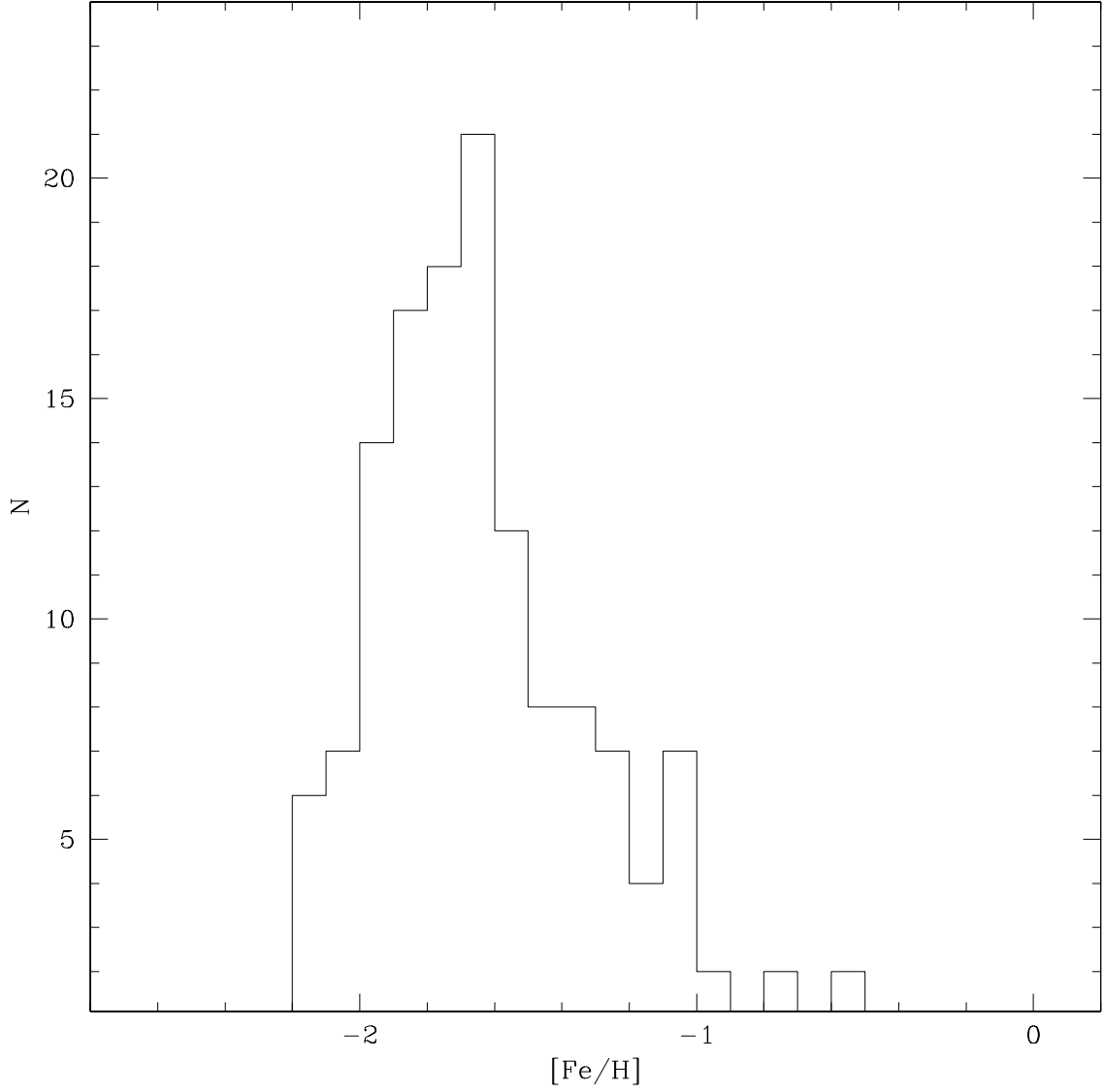}
\caption{The derived metallicity distribution for the 152 cluster member stars
in the $WFI~sample$.} 
\label{histo}
\end{figure}

\clearpage

\clearpage

\clearpage

\clearpage

\clearpage

\clearpage

\clearpage

\clearpage
 
\clearpage
 
\clearpage

\begin{deluxetable}{lcccc}
\tablewidth{0pt}
\tablecaption{Ca II triplet lines and continuum bands}
\tablehead{
\colhead{Feature}      & \colhead{Line}                &
\colhead{Line}         & \colhead{Blue continuum}      &
\colhead{Red continuum}\\
\colhead{name}         & \colhead{center}              &
\colhead{band}         & \colhead{band}                &
\colhead{band}\\
\colhead{}             & \colhead{($\AA$)}             &
\colhead{($\AA$)}      & \colhead{($\AA$)}             &
\colhead{($\AA$)}}
\startdata
$\lambda_{8498}$ & 8498.1 & 8490 -- 8506 & 8346 -- 8489 & 8563 -- 8642\\
$\lambda_{8542}$ & 8542.3 & 8532 -- 8552 & 8346 -- 8489 & 8563 -- 8642\\
$\lambda_{8662}$ & 8662.4 & 8653 -- 8671 & 8563 -- 8642 & 8697 -- 8754\\
\enddata
\end{deluxetable}

\clearpage

\begin{deluxetable}{lcccr}
\tablewidth{0pt}
\tablecaption{The populations of $\omega$ Cen}
\tablehead{
\colhead{Population}     & \colhead{[Fe/H]} &
\colhead{[$\alpha$/Fe]}  & \colhead{Y}      &
\colhead{$\Delta$ Age (Gyr)}}
\startdata
 MP         & -1.7   & 0.3           & 0.246  & 0           \\
	    &        &               &        &\\
 MInt2	    & -1.3   & 0.3           & 0.246  & 0           \\
 MInt2      & -1.3   & 0.3           & 0.30   & 0           \\
 MInt2      & -1.3   & 0.3           & 0.35   & 0           \\
 	    &	     &		     &	      &\\
 MInt3	    & -1.0   & 0.3           & 0.246  & 1           \\
 MInt3      & -1.0   & 0.3           & 0.30   & 1           \\
 MInt3      & -1.0   & 0.3           & 0.35   & 0           \\
            &        &               &        &\\
 a          & -0.6   & 0.1           & 0.246  & $>$1        \\
 a          & -0.6   & 0.1           & 0.30   & 0 $\div$ 1  \\
 a          & -0.6   & 0.1           & 0.35   & 0 $\div$ 1  \\
 a          & -0.6   & 0.1           & 0.40   & 0           \\
\enddata
\end{deluxetable}

\end{document}